\documentclass{elsart}
\usepackage{amssymb}
\newcommand{\sect}[1]{\setcounter{equation}{0}\section{#1}}
\renewcommand{\theequation}{\arabic{section}.\arabic{equation}}
\newcommand{\bfm}[1]{\mbox{\boldmath${#1}$}}
\newcommand{\app}{\setcounter{section}{0}
\setcounter{equation}{0}
\renewcommand{\thesection}{Appendix}\renewcommand{\theequation}{\Alph{section}.
\arabic{equation}}}
\usepackage{latexsym}
\usepackage{graphics}
\begin{document}
\begin{frontmatter}
\title{Equivalence among different formalisms in the Tsallis entropy framework}
\author{A.M. Scarfone}
\address{Istituto Nazionale di
Fisica della Materia (CNR-INFM) and Dipartimento di Fisica,
Politecnico di Torino, Corso Duca degli Abruzzi 24, 10129 Torino,
Italy.}

\author{T. Wada}
\address{Department of Electrical and
Electronic Engineering,\\ Ibaraki University, Hitachi, Ibaraki,
316-8511, Japan.}

\date{\today}
\begin{abstract}
In a recent paper [Phys. Lett. A {\bf335}, 351 (2005)] the authors
discussed the equivalence among the various probability distribution
functions of a system in equilibrium in the Tsallis entropy
framework. In the present letter we extend these results to a system
which is out of equilibrium and evolves to a stationary state
according to a nonlinear Fokker-Planck equation. By means of
time-scale conversion, it is shown that there exists a
``correspondence'' among the self-similar solutions of the nonlinear
Fokker-Planck equations associated with the different Tsallis
formalisms. The time-scale conversion is related to the
corresponding Lyapunov functions of the respective nonlinear
Fokker-Planck equations.
\end{abstract}
\end{frontmatter}
\maketitle

\sect{Introduction}

In the last decades some generalized entropies, different from the
Boltzmann-Gibbs one, have been proposed in literature to study the
statistical mechanics properties of complex systems with asymptotic
free-scale behavior. Among the various proposals, the Tsallis' entropy
\cite{Tsallis}
\begin{equation}
S_q[p]={1\over1-q}\int\Big(p(v)^q-p(v)\Big)\,dv \ ,
\end{equation}
is a paradigm.\\ After its introduction in 1988, it has been
fruitfully employed in the investigation of statistical systems
characterized by long-rang interactions and time-persistent memory
effects. Among the many, we quote astrophysical and cosmological
systems, nuclear systems, low-dimensional chaotic systems,
self-organized critical systems, turbulence. It is also widely
employed in geological, biological and social-economical sciences
(see \cite{Tsallis9} for a
currently update bibliography).\\
As observed in \cite{Sisto}, the original proposal \cite{Tsallis}
introduced the Lagrange multiplier associated with the linear mean
energy in an unsatisfied manner so that some important properties
necessary for a well stated theory were lacked. This insight induced
in \cite{Sisto,Bashkirov} to reformulate the original formalism
(hereinafter we call it the modified 1st-formalism) based on
the definition of the linear mean value given by
\begin{equation}
A^{(1)}=\int A(v)\,p^{(1)}(v)\,dv \ . \label{lin}
\end{equation}
In \cite{Tsallis1,Tsallis2} alternative
definitions of the mean value were introduced, and the authors
proposed the 2nd-formalism based on the un-normalized $q$-mean value
\begin{equation}
A^{(2)}=\int A(v)\,\Big(p^{(2)}(v)\Big)^q\,dv \ ,
\end{equation}
which unfortunately has the drawback that $\langle
1\rangle^{(2)}\not=1$ and the 3rd-formalism based on the escort
probability mean value
\begin{equation}
A^{(3)}={\int
A(v)\,\Big(p^{(3)}(v)\Big)^q\,dv\over\int
\Big(p^{(3)}(v)\Big)^q\,dv} \ . \label{esc}
\end{equation}
More recently, in \cite{Martinez,Lenzi} the
optimal Lagrange multiplier formalism (OLM-formalism) were introduced, where for the
same physical input of the 3rd-formalism, the constraint in the MaxEnt
procedure is written as the {\em centered}-form
\begin{equation}
{\int A(v)\Big(p^{(\rm OLM)}(v)\Big)^q\,dv-A^{(\rm OLM)}
\int \Big(p^{(\rm OLM)}(v)\Big)^q\,dv} = 0.
\end{equation}
The 1st- and 3rd-formalisms have the disadvantage of dealing with
self-referential expressions for the distribution functions which
originate computational difficulties in some cases. Differently, the
2nd- and the OLM-formalisms produce distribution functions that are
not self-referential though the mean values are still
self-referential (see
\cite{Ferri1} for a comparison between these two formalisms).\\
Throughout all these alternative formulations, in
\cite{Scarfone1,Wada1}, another formalism based on the entropy
$S_{2-q}[p]$ and the linear averages (hereinafter $(2-q)$-formalism)
has been developed. In particular, in \cite{Wada3} it has been shown
that the $(2-q)$-formalism produces a non self-referential
expression both in the distribution and in the mean values. (For a comparison of some physically relevant quantities in the different formalisms see Appendix A). \\ In
\cite{Wada2}, the entropy $S_{2-q}[p]$ and its equilibrium
distribution have been employed to show the equivalence among the
different formulations of the Tsallis theory. The same
$(2-q)$-formalism plays an important role in the discussion of the
present work. For this reason, and also to fix the notations used in
this paper, we shortly collect hereinafter the main results obtained in
\cite{Wada2}.\\
According to the following relations
\begin{eqnarray}
&&\beta^{(1)}=(2-q)\,\beta^{(1)}_{2-q}\,\Big(\bar
Z_{2-q}^{(1)}\Big)^{q-1} \ ,\label{b1}\\
&&\gamma^{(1)}={1\over1-q}-{2-q\over1-q}\,\Big(\bar
Z_{2-q}^{(1)}\Big)^{q-1}-(2-q)\,\beta_{2-q}^{(1)}\,\Big(\bar
Z_{2-q}^{(1)}\Big)^{q-1}\,U^{(1)}_q \ ,\label{g1}
\end{eqnarray}
the equilibrium distribution obtained in the $(2-q)$-formalism
\begin{equation}
p(v)=\alpha\,\exp_q\Big(-\gamma^{(1)}-\beta^{(1)}\,\epsilon(v)\Big)
\ ,\label{pdf0}
\end{equation}
with $\alpha=(2-q)^{1/(q-1)}$ and $\exp_q(x)=[1+(1-q)\,x]^{1/(1-q)}$ is the $q$-deformed exponential, can be rewritten in the form
\begin{equation}
p^{(1)}(v)={1\over\bar
Z_{2-q}^{(1)}}\,\exp_q\Big(-\beta_{2-q}^{(1)}\,\Big(\epsilon(v)-U^{(1)}_q\Big)\Big)
\ ,\label{pdf1}
\end{equation}
with $(\bar
Z_{2-q}^{(1)})^{q-1}=\int(p^{(1)}(v))^{2-q}\,dv$, where, by replacing $2-q$ with $q$, we obtain the probability distribution function of the modified
1st-formalism \cite{Bashkirov}. In Eq. (\ref{pdf0}), $\gamma^{(1)}$ and $\beta^{(1)}$ are the Lagrange multipliers related, respectively, to the normalization $1=\int p(v)\,dv$ and to the linear mean energy
$U_q=\int\epsilon(v)\,p(v)\,dv$ in the ($2-q$)-formalism, as well as $\beta^{(1)}_{2-q}$ is related to the Lagrange
multiplier associated to the linear mean energy
$U^{(1)}_q=\int\epsilon(v)\,p^{(1)}(v)\,dv$ in the 1st-formalism.\\ Equivalently, the
same expression (\ref{pdf0}) can be transformed in
\begin{equation}
p^{(3)}(v)={1\over\bar
Z_q^{(3)}}\,\exp_q\Big(-\beta_q^{(3)}\,\Big(\epsilon(v)-U^{(3)}_q\Big)\Big)
\ ,\label{pdf3}
\end{equation}
corresponding to the distribution function derived in the
3rd-formalism \cite{Tsallis2}, where
$\beta^{(3)}=\beta^{(3)}_q\,(\bar Z_q^{(3)})^{1-q}$ is the Lagrange
multiplier of the escort mean energy
$U^{(3)}_q={\int\epsilon(v)\,(p^{(3)}(v))^q\,dv\over
\int(p^{(3)}(v))^q\,dv}$ and $(\bar Z_q^{(3)})^{1-q}=\int
(p^{(3)}(v))^q\,dv$.\\
The expressions (\ref{pdf1}) and (\ref{pdf3}) are equivalents, in
the sense that $p^{(1)}(v)\equiv p^{(3)}(v)$, if the conditions
\begin{equation}
\beta^{(1)}=(2-q)\,\beta^{(3)} \,\left(\bar
Z_q^{(3)}\right)^{2(q-1)} \ ,\label{beta}
\end{equation}
and
\begin{equation}
U^{(3)}_q=U^{(1)}_q-{2-q\over(1-q)\,\beta^{(1)}}\,\left[\Big(\bar
Z_q^{(3)}\Big)^{q-1}- \Big(\bar Z_{2-q}^{(1)}\Big)^{q-1}\right] \
,\label{u}
\end{equation}
are fulfilled. (See also \cite{Ferri2} for an alternative discussion
on the equivalence among these different formalisms).\\
In a previous tentative to establish a correspondence among the
various formulations described above, it has been introduced the
escort entropy \cite{Tsallis2,Tsallis3}
\begin{equation}
S_q^{\rm (E)}[p]={1\over1-q}\left\{\frac{\int
p^{\rm(E)}(v,\,t)\,dv}{\left[\int
\Big(p^{\rm(E)}(v,\,t)\Big)^{1/q}\,dv\right]^q}-1\right\} \
.\label{escort}
\end{equation}
This entropy is derived from $S_q[p]$ by using the transformation
\begin{equation}
p^{(3)}(v)={\Big(p^{\rm(E)}(v)\Big)^{1/q}\over\int\Big(p^{\rm(E)}(v)\Big)^{1/q}\,dv}
\ ,
\end{equation}
where distributions $p^{(3)}(v)$ and $p^{\rm(E)}(v)$ came,
respectively, from the entropy $S_q[p]$ constrained by the escort
mean value $U^{(3)}_q$ and from the entropy $S^{\rm (E)}_{\tilde
q}(p)$, with $\tilde q=1/q$, constrained by linear mean value
$U^{\rm (E)}_{\tilde q}=\int\epsilon(v)\,p^{\rm (E)}(v)\,dv$.\\ In
the following, we are concerned with self-similar solutions of a
nonlinear Fokker-Planck equation (NFPE) derived from the
$S_{2-q}[p]$ entropy with a linear drift. This kinetic equation, and
some related versions, are widely considered in the literature
\cite{Stephenson,Pascal,Pascal1} to study anomalous diffusion
processes like fractal media diffusion or diffusion in a
non-Newtonian fluid. Self-similar solutions corresponding to these
NFPEs were also investigated in \cite{Lapenta,Compte2,Compte1}
because their relevance in the description
of several phenomenologies in condensed matter physics.\\
The purpose of the present work is twofold.\\
Firstly, we extend the discussion held in \cite{Wada2} to the case
of a system out of equilibrium. This is performed in the next
section 2, where we show that the self-similar solutions derived in
the $(2-q)$-formalism are equivalent to those of the same kinetic
equations obtained by employing the modified
1st-formalism and 3rd-formalism.\\
Secondly, in section 3, we show that, by means of a time-scale
conversion, there is a ``correspondence'' among the (self-similar)
solutions of the NFPE derived in the $(2-q)$-formalism and  the
(self-similar) solutions of the NFPE derived in the
escort-formalisms. As a by product, it is shown that there exists a
``correspondence'' with the (self-similar) solutions of the NFPE
derived in the 3rd-formalism with a time dependent diffusion
coefficient. To establish a relationship among different NFPEs is
useful not only in the framework of the statistical mechanics based
on the Tsallis entropy but in general with the purpose to classify
the various kinetic equations in equivalence class such that any
equation belonging to the same class describes substantially the
same physical process. The conclusions are given in section 4 whilst
a comparison among the different formalisms discussed in this paper is reported in the Appendix A.


\sect{Nonlinear Fokker-Planck equation in the
$\bfm{(2-q)}$-formalism}

In order to describe our problem we need to relate the production of
entropy to a Fokker-Planck equation describing the kinetics of the
system. This issue has been discussed in literature by several
authors, both in a very general nonlinear kinetic framework
\cite{Frank1,Frank2,Kaniadakis} and, more specifically, in the
Tsallis
entropy framework \cite{Compte1,Plastino,Shiino1}.\\
Quite generally, the NFPE associated with a given entropy $S[p]$ can
be obtained from the relation
\begin{equation}
\frac{\partial}{\partial
t}\,p(v,\,t)=-{\partial\over\partial\,v}\,J(v,\,t) \ ,\label{FP}
\end{equation}
and assuming for the expression of the nonlinear current
\begin{equation}
J(v,\,t)=-p(v,\,t)\,{\partial\over\partial\,v}\,\left(\frac{\delta
{\mathcal L}}{\delta p}\right) \ .\label{j}
\end{equation}
The quantity $\partial\,(\delta{\mathcal L}/\delta p)/\partial\,v$
is the thermodynamic force, where ${\mathcal L}[p]$ is the Lyapunov
functional for the given problem \cite{Frank4}.\\
By posing
\begin{equation}
{\mathcal L}[p] = {\mathcal L}^{(1)}_q[p] \equiv
U^{(1)}_q-D^{(1)}\,S_{2-q}[p] \ ,\label{Lyapunov1}
\end{equation}
where
\begin{equation}
S_{2-q}[p]=\int\frac{\Big(p(v,\,t)\Big)^{2-q}-p(v,\,t)}{q-1}\,dv \ ,
\end{equation}
is the Tsallis entropy in the $(2-q)$-formalism and $D^{(1)}$ is the
constant diffusion coefficient, we obtain the following NFPE
\begin{equation}
\frac{\partial}{\partial t}
p(v,\,t)={\partial\over\partial\,v}\,\left[-h(v)\,p(v,\,t)+D^{(1)}\,
{\partial\over\partial\,v}\,\Big(p(v,\,t)\Big)^{2-q}\right] \
,\label{FP1}
\end{equation}
where
\begin{equation}
h(v)=-{\partial\,\epsilon(v)\over\partial\,v} \ ,
\end{equation}
is the drift coefficient. Equation (\ref{FP1}) differs, in $q\to
2-q$, from the kinetics equation derived in
\cite{Compte2,Compte1,Frank1} from entropy $S_q[p]$. The same Eq.
(\ref{FP1}) has been first proposed in \cite{Plastino} and after
reconsidered by different authors \cite{Borland1,Drazer,Borland2,Tsallis4}.\\ In the following, we assume
$\epsilon(v)={1\over2}\,m\,v^2$ with $m>0$, which deals with a
specific but very common expression for the drift coefficient,
corresponding to the Uhlenbeck-Ornstein process in the $q\to1$
limit. Without loss of generality, we can pose $m=1$. With this
choice for $\epsilon(v)$, Eq. (\ref{FP1}) admits self-similar
solutions \cite{Lapenta,Plastino,Tsallis5}. In the $(2-q)$-formalism
such solutions are obtained by introducing the ansatz
\begin{equation}
p(v,\,t)=\alpha\,\exp_q\Bigg(-\gamma^{(1)}(t)-\beta^{(1)}(t)\,{1\over2}\,v^2\Bigg)
\ .\label{nsr}
\end{equation}
In this way, we derive the following system of differential
equations for the unknown quantities $\gamma^{(1)}(t)$ and
$\beta^{(1)}(t)$:
\begin{eqnarray}
&&\frac{d}{d\,t}\,\gamma^{(1)}(t)+
\Big[1-(1-q)\,\gamma^{(1)}(t)\Big]\,\left(1-D^{(1)}\,\beta^{(1)}(t)\right)
=0 , \label{s1}\\
&&\frac{d}{d\,t}\,\beta^{(1)}(t)
-(3-q)\,\beta^{(1)}(t)\,\left(1-D^{(1)}\,\beta^{(1)}(t)\right)=0 \
.\label{s2}
\end{eqnarray}
Moreover, accounting for the constraints on the normalization and
the linear mean energy, which give the
relations\footnote{Hereinafter, for the sake of simplicity, we
assume $1<q<{5\over3}$ to guarantee the convergence of all integrals
on the full real axe $(-\infty,\,+\infty)$. The results can be
straightforward generalized to $0<q<1$, accounting for the cut-off
condition on the integration interval.}
\begin{equation}
\gamma^{(1)}(t)=\ln_{2-q}\left(\alpha_1\,\sqrt{U^{(1)}_q(t)}\right)
\
,\hspace{10mm}\beta^{(1)}(t)=\alpha_2\,\left(U^{(1)}_q(t)\right)^{\frac{q-3}{2}}
\ ,\label{def}
\end{equation}
with
\begin{equation}
\alpha_1=\alpha\,\sqrt{2\,\pi\,{5-3\,q\over
q-1}}\,{\Gamma\left({1\over
q-1}-{1\over2}\right)\over\Gamma\left({1\over q-1}\right)} \
,\hspace{10mm} \alpha_2={\alpha_1^{q-1}\over5-3\,q} \ ,
\end{equation}
from Eqs. (\ref{s1}) or (\ref{s2}) we obtain the further
differential equation
\begin{equation}
\frac{\partial}{\partial\,t}\,\ln U^{(1)}_q(t)+2\,\left[1-\alpha_2\,
D^{(1)}\,\Big(U^{(1)}_q(t)\Big)^{\frac{q-3}{2}}\right]=0 \
,\label{eve}
\end{equation}
describing the time evolution of the mean energy.\\
Solutions of Eqs. (\ref{s1}), (\ref{s2}) and (\ref{eve}) are given
by
\begin{eqnarray}
&&\gamma^{(1)}(t)=\gamma_0-\Big[1-(1-q)\,
\gamma_0\Big]\,\ln_q\left({\beta^{(1)}(t)\over\beta_0}\right)^{1\over3-q} \ ,\label{sol1a}\\
&&\beta^{(1)}(t)=\beta_0\Bigg[\Big(1-\beta_0\,D^{(1)}\Big)\,e^{(q-3)\,t}
+\beta_0\,D^{(1)}\Bigg]^{-1} \ ,\label{sol1b}\\
&&U^{(1)}_q(t)=U_0\,\Bigg[\Big(1-\mu_0\,D^{(1)}\Big)\,e^{(q-3)\,t}
+\mu_0\,D^{(1)}\Bigg]^{\frac{2}{3-q}} \ ,\label{sol1c}
\end{eqnarray}
with $\mu_0\equiv\beta_0=\alpha_2\,U_0^{\frac{q-3}{2}}$,
$\gamma_0=\ln_{2-q}(\alpha_1\,\sqrt{U_0})$ and $U_0=U^{(1)}_q(0)$.
At equilibrium, $t\to t_{\rm e}=+\infty$, from Eq. (\ref{sol1b}) we
obtain the relation $\beta^{(1)}_{\rm e}\,D^{(1)}=1$, where
$\beta^{(1)}_{\rm e}\equiv\beta^{(1)}(t_{\rm e})$, which means, as
already known \cite{Frank2,Shiino1}, that the quantity $1/D^{(1)}$
can be identified with the Lagrange multiplier related to
the mean energy $U_q^{(1)}$.\\
Differently, by using in Eq. (\ref{FP1}) the ansatz
\begin{equation}
p^{(1)}(v,\,t)={1\over\bar
Z_{2-q}^{(1)}(t)}\,\exp_q\Bigg(-\beta_{2-q}^{(1)}(t)\,\Bigg(
{1\over2}\,v^2-U^{(1)}_q(t)\Bigg)\Bigg) \ ,\label{1st}
\end{equation}
for the modified 1st-formalism, we obtain the following set of
equations
\begin{equation}
\frac{d}{d\,t}\ln X^{(1)}_q(t)+2\,f_q^{(1)}(t)=0 \ ,\label{eq0}
\end{equation}
where $X^{(1)}_q(t)$ means $\Big(\bar
Z_{2-q}^{(1)}(t)\Big)^2,\,1/\beta_{2-q}^{(1)}(t)$ and
$U^{(1)}_q(t)$, respectively, with
\begin{equation}
f^{(1)}_q(t)=1-D^{(1)}\,\beta^{(1)}(t) \ ,\label{f1}
\end{equation}
and we posed $\beta^{(1)}(t)=(2-q)\,\beta_{2-q}^{(1)}(t)\,(\bar
Z_{2-q}^{(1)}(t))^{q-1}$, according to Eq. (\ref{b1}).\\ By means of
the relations (\ref{b1}) and (\ref{g1}) the system (\ref{eq0}) is
equivalent to the system (\ref{s1}), (\ref{s2}) and (\ref{eve}).
Correspondingly, its solutions
\begin{equation}
\Big(\bar Z_{2-q}^{(1)}(t)\Big)^2\,\beta_{2-q}^{(1)}(t)=
\frac{\Big(z_{2-q}^{(1)}\Big)^2}{2-q} \ ,\hspace{10mm}
U^{(1)}_q(t)\,\beta_{2-q}^{(1)}(t)=\frac{1}{2\,(2-q)} \
,\label{sol22}
\end{equation}
with
\begin{equation}
z_{2-q}^{(1)}=\alpha\,\sqrt{2\,\pi\over
q-1}\left(\frac{5-3\,q}{2}\right)^{3-q\over2(1-q)}\,{\Gamma\left({1\over
q-1}-{1\over2}\right)\over\Gamma\left({1\over q-1}\right)} \
,\label{zq1}
\end{equation}
are equivalent to the solutions (\ref{sol1a})-(\ref{sol1c}), whilst
the time evolution of $U^{(1)}_q(t)$ is still given by Eq.
(\ref{sol1c}).\\
In a similar way, with the ansatz
\begin{equation}
p^{(3)}(v,\,t)={1\over\bar
Z_q^{(3)}(t)}\,\exp_q\Bigg(-\beta_q^{(3)}(t)\,\Bigg({1\over2}\,v^2-U^{(3)}_q(t)\Bigg)\Bigg)
\ ,\label{3st}
\end{equation}
from Eq. (\ref{FP1}) we obtain the system of the differential
equations
\begin{equation}
\frac{d}{d\,t}\ln X^{(3)}_q(t)+2\,f_q^{(3)}(t)=0 \ ,\label{eq1}
\end{equation}
where $X^{(3)}_q(t)$ means $\Big(\bar
Z_q^{(3)}(t)\Big)^2,\,1/\beta_q^{(3)}(t)$ and $U_q^{(3)}(t)$,
respectively, and
\begin{equation}
f^{(3)}_q(t)=1-D^{(1)}\,(2-q)\,\beta^{(3)}(t)\,\left(\bar
Z_q^{(3)}(t)\right)^{2\,(q-1)} \ ,
\end{equation}
with $\beta^{(3)}(t)=\beta_q^{(3)}(t)\,(\bar Z_q^{(3)}(t))^{1-q}$.
We note that at equilibrium ($t=t_{\rm e}$), $\beta^{(3)}(t_{\rm
e})$ can be identified with the Lagrange multiplier of the escort
mean energy
$U_q^{(3)}(t_{\rm e})$.\\
By using Eqs. (\ref{beta}) and (\ref{u}), it is easy to verify that
the system (\ref{eq1}) can be transformed in the system (\ref{eq0})
and therefore, it is also equivalent to the system (\ref{s1}),
(\ref{s2}) and (\ref{eve}).\\ The solutions of Eqs. (\ref{eq1}) are
given by
\begin{equation}
\Big(\bar
Z_q^{(3)}(t)\Big)^2\,\beta_q^{(3)}(t)=\Big(z_q^{(3)}\Big)^2 \
,\hspace{10mm}U^{(3)}_q(t)\,\beta_q^{(3)}(t)={1\over2} \
,\label{sol3}
\end{equation}
where
\begin{equation}
z_q^{(3)}=\sqrt{2\,\pi\over
q-1}\left(\frac{3-q}{2}\right)^{3-q\over2(1-q)}\,{\Gamma\left({1\over
q-1}-{1\over2}\right)\over\Gamma\left({1\over q-1}\right)} \
,\label{zq3}
\end{equation}
whilst the solution for the mean energy $U^{(3)}_q(t)$ has the same
expression given in Eq. (\ref{sol1c}), with $U_q^{(1)}\to U_q^{(3)}$
and $D^{(1)}\to D^{(3)}$, being now
$\mu_0=\alpha_2\left(\frac{3-q}{5-3\,q}\right)^{\frac{q-3}{2}}U_0^{\frac{q-3}{2}}$
and $U_0=U^{(3)}_q(0)$.

\sect{``Correspondence'' between the modified 1st-formalism and the
3rd-formalism}

In this section, we study a different kind of relationship, which we
call ``correspondence'', between
the modified 1st-formalism and 3rd-formalism\\
To start with, let us observe that by posing $q=2-1/q^\prime$
through the definitions (\ref{zq1}) and (\ref{zq3}) we verify that
\begin{equation}
z_{2-q}^{(1)}=z_{q'}^{(3)} \ .
\end{equation}
On the other hand, by assuming the same initial values for the both
mean energy
\begin{equation}
U^{(1)}_q(0)=U^{(3)}_{q^\prime}(0) \ ,\label{cond}
\end{equation}
and by comparing Eqs. (\ref{sol22}) and (\ref{sol3}) we get
\begin{equation}
\beta_{2-q}^{(1)}(0)=q'\,\beta_{q^\prime}^{(3)}(0) \ ,\hspace{20mm}
\bar Z_{2-q}^{(1)}(0)=\bar Z_{q^\prime}^{(3)}(0) \ .\label{zz}
\end{equation}
As a consequence, the following relation
\begin{equation}
p_q^{(1)}(v,\,0)=\left(\bar
Z_{q^\prime}^{(3)}(0)\right)^{q^\prime-1}
\,\Big(p_{q^\prime}^{(3)}(v,\,0)\Big)^{q^\prime}\equiv
p_{q'}^{\rm(E)}(v,\,0) \ ,\label{p1-p3}
\end{equation}
and its inverse
\begin{equation}
p_{q'}^{(3)}(v,\,0)=\left(\bar Z_{2-q}^{(1)}(0)\right)^{1-q}
\,\Big(p_q^{(1)}(v,\,0)\Big)^{2-q} \ ,\label{p3-p1}
\end{equation}
hold, where we denoted\footnote{From now on, for the sake of
clarity, we indicate explicitly the dependence on the parameter $q$
($q'$ respectively) in the distribution.} with
$p^{\rm(E)}_{q'}(v,\,0)$ the escort distribution obtained from the
optimization problem with the entropy $S^{\rm E}_{q'}[p]$ and the
corresponding linear mean energy $U_{q'}^{\rm(E)}$.
\\
The consistency of Eqs. (\ref{p1-p3}) and (\ref{p3-p1}) can be
easily verified as follows. First, by integrating Eq. (\ref{p1-p3})
over $v$ and utilizing the normalization of $p_q^{(1)}(v,\,0)$ and
$p_{q'}^{(3)}(v,\,0)$, it reduces to the definition of the partition
functions $\bar Z_{q^\prime}^{(3)}(0)$ and $\bar Z_{2-q}^{(1)}(0)$.
Second, by multiplying Eq. (\ref{p1-p3}) by $\epsilon(v)$ and
integrating over $v$, we can verify that
\begin{equation}
U^{(1)}_q(0)=\int \epsilon(v)\,p_q^{(1)}(v,\,0)\,dv=
\frac{\int\epsilon(v)\,\Big(p_{q'}^{(3)}(v,\,0)\Big)^{q'}\,dv}
{\int\Big(p_{q'}^{(3)}(v,\,0)\Big)^{q'}\,dv}=U^{(3)}_{q'}(0) \
,\label{eqe}
\end{equation}
which is the initial condition (\ref{cond}).\\
All of these results are already known for the equilibrium system.
Notwithstanding, let us now pose the following question: Does the
``correspondence'' established through Eqs. (\ref{p1-p3}) and
(\ref{p3-p1}) still hold, in some sense, when the
system evolves toward the equilibrium according to a suitable NFPEs?\\
For reasons which will be clarified through this section, it is more
appropriate to start by studying the ``correspondence'' among
$p^{(1)}_q(v,\,t)$ and $p_{q'}^{\rm(E)}(v,\,t)$ and only
successively recover the ``correspondence'' with
$p^{(3)}_{q'}(v,\,t)$ by means of Eq. (\ref{p1-p3}). In fact, by
employing the escort-formalism we get the advantage to deal with a
linear expression for the mean energy $U^{\rm(E)}_{q'}(t)\equiv
U^{(3)}_{q'}(t)=\int\epsilon(v)\,p_{q'}^{\rm(E)}(v,\,t)\,dv$. In
this way, by posing
\begin{equation}
{\mathcal L} = {\mathcal L}^{\rm(E)}_{q'}[p] \equiv
U^{\rm(E)}_{q'}(t)-D^{\rm (E)}\,S_{q'}^{\rm (E)}[p] \
,\label{Lyapunov3}
\end{equation}
where $S^{\rm (E)}_q[p]$ is the escort entropy (\ref{escort}), we derive
the following kinetic equation:
\begin{eqnarray}
\nonumber \frac{\partial}{\partial t}
p_{q'}^{\rm(E)}(v,\,t)=\frac{\partial}{\partial
v}\left[v\,p_{q'}^{\rm(E)}(v,\,t)+D^{\rm (E)}\,\left(\bar
Z_{q'}^{\rm (E)}(t)\right)^{{1\over
q'}-q'}\,\frac{\partial}{\partial
v}\Big(p_{q'}^{\rm(E)}(v,\,t)\Big)^{1\over q'}\right] \
,\\\label{FPNe}
\end{eqnarray}
where $\bar Z_{q'}^{(E)}(t)=\bar Z_{q'}^{(3)}(t)=(\int
p_{q'}^{\rm(E)}(v,\,t)^{1/q'}\,dv)^{q'/(q'-1)}$ and $D^{(E)}$ is the
constant diffusion coefficient for the given problem. This equation, for $q=1/q^\prime$, becomes equivalent to the NFPE obtained previously in \cite{Shiino1}.\\
Let us observe that Eqs. (\ref{FP}) and (\ref{j}) provide a general mechanism to construct NFPEs starting from a give Lyapunov functional ${\mathcal L}[p]$. This method works also when the mean energy term in the Lyapunov functional has a nonlinear dependence on the probability distribution functions like, for instance, in the case of $U_q^{(3)}$. Notwithstanding, in these cases the resulting nonlinear current $J(v,\,t)$ will contain, among to a nonlinear diffusion term, also a nonlinear drift term. This fact makes problematic the comparison between the different formalisms which is the main purpose of the present work. A possible way to avoid such complication is actually furnished by means of the escort formalism centered on the linear definition for the mean energy $U_q^{\rm(E)}$.\\
Come back to the original question, self-similar solutions of Eq. (\ref{FPNe}) can be obtained by posing
\begin{equation}
p_{q'}^{\rm(E)}(v,\,t)={1\over \bar
Z_{q'}^{(E)}(t)}\Bigg[1-(1-q')\,\beta_{q'}^{\rm(E)}(t)
\Bigg({1\over2}\,v^2-U^{\rm (E)}_{q'}(t)\Bigg)\Bigg]^{q'\over1-q'} \
,\label{est}
\end{equation}
where $\beta_{q'}^{\rm(E)}(t)=\beta^{\rm(E)}(t)\,\left(\bar
Z_{q'}^{\rm(E)}(t)\right)^{{1\over q'}-1}$ and at equilibrium
$\beta^{\rm(E)}(t_{\rm e})$ is the Lagrange multiplier related to
the escort mean energy $U^{\rm(E)}_{q'}$.\\ We get the following
system of equations
\begin{equation}
\frac{\partial}{\partial t}\ln
X_{q'}^{\rm(E)}(t)+2\,f^{\rm(E)}_{q'}(t)=0 \ ,\label{eq2}
\end{equation}
where $X^{\rm(E)}_q(t)$ means $\Big(\bar
Z_q^{\rm(E)}(t)\Big)^2,\,1/\beta_q^{\rm(E)}(t)$ and $U_q^{\rm
(E)}(t)$, respectively, and
\begin{equation}
f^{\rm(E)}_q(t)=1-D^{\rm(E)}\,\beta^{\rm(E)}(t)\,\left(\bar
Z_q^{\rm(E)}(t)\right)^{\frac{1}{q}-q} \ .\label{fe}
\end{equation}
The solution of the system (\ref{eq2}) can be easily written in the
form
\begin{equation}
\left(\bar
Z_{q'}^{\rm(E)}(t)\right)^2\,\beta_{q'}^{\rm(E)}(t)=\Big(z_{q'}^{(3)}\Big)^2
\ ,\hspace{10mm}U_{q'}^{\rm(E)}(t)\,\beta_{q'}^{\rm(E)}(t)={1\over2}
\ ,\label{soln}
\end{equation}
and
\begin{equation}
U^{\rm(E)}_{q'}(t)=U_0\left[\left(1-D^{\rm(E)}\,\mu_0^{\rm(E)}\right)\,
e^{-(1+q')\,t}+D^{\rm(E)}\,\mu_0^{\rm(E)}\right]^{\frac{2}{1+q'}}
\end{equation}
with
$\mu_0^{\rm(E)}=2^{-\frac{1+q'}{2}}\,(z_{q'}^{(3)})^{1-q'}\,U_0^{\frac{1-q'}{2}}$.

Let us now introduce the two different time-scales ($t$ and $t'$)
according to the relation
\begin{equation}
\frac{dt'}{dt}=\frac{f^{(1)}_q(t)} {f^{\rm(E)}_{q'}(t')} \
.\label{tt}
\end{equation}
Accounting for the expressions (\ref{f1}) and (\ref{fe}), Eq.
(\ref{tt}) can be integrated in
\begin{eqnarray}
\nonumber\hspace{-5mm}
t'=t_0-{1\over1+q'}\,\ln\left\{\left[\left(1-D^{(1)}\,\mu_0^{(1)}\right)\,
e^{-\big({1\over
q'}+1\big)\,t}+D^{(1)}\,\mu_0^{(1)}\right]^{q'}-D^{\rm(E)}\,\mu_0^{\rm(E)}\right\}
\ ,\\ \label{time}
\end{eqnarray}
where
\begin{equation}
t_0={1\over1+q'}\ln(1-D^{\rm(E)}\,\mu_0^{\rm(E)}) \ .
\end{equation}
By converting the time-scale $t$ into $t'$, we can easily verify
that the set of equations for the modified 1st formalism are
transformed to those for the escort formalism
\begin{equation}
\frac{d}{d\, t}\ln X_q^{(1)}(t)+f^{(1)}_q(t)=0 \
,\hspace{5mm}\Rightarrow\hspace{5mm}\frac{d}{d\, t'}\ln
X_{q'}^{\rm(E)}(t')+f^{\rm(E)}_{q'}(t')=0 \ ,
\end{equation}
which implies the following ``correspondence''
\begin{equation}
X_q^{(1)}(t)=X_{q'}^{\rm(E)}(t') \ .\label{en}
\end{equation}
Since the quantities $X_q^{(1)}(t)$ and $X_{q'}^{\rm(E)}(t')$
determine completely the shape of the distribution
$p^{(1)}_q(v,\,t)$ and that of $p_{q'}^{\rm (E)}(v,\,t')$,
respectively, we conclude that the ``correspondence'' stated in Eq.
(\ref{p1-p3}) at the initial time actually holds for any later time,
i.e.
\begin{equation}
p_q^{(1)}(v,\,t)=p_{q'}^{\rm(E)}(v,\,t') \ .\label{pP}
\end{equation}
Finally, taking into account the expressions (\ref{1st}) and
(\ref{est}), from the relation (\ref{pP}) we obtain
\begin{eqnarray}
\nonumber&&\hspace{-8mm}\frac{\partial}{\partial t}
p_q^{(1)}(v,\,t)=\frac{\partial}{\partial
v}\left[v\,p_q^{(1)}(v,\,t)+D^{(1)}\,\frac{\partial}{\partial
v}\left(p_q^{(1)}(v,\,t)\right)^{2-q}\right]\hspace{10mm}\Rightarrow
\\
\nonumber &&\hspace{-8mm}\frac{\partial}{\partial t'}
p_{q'}^{\rm(E)}(v,\,t')=\frac{\partial}{\partial
v}\left[v\,p_{q'}^{\rm(E)}(v,\,t')+D^{\rm(E)}\,\left(\bar
Z_{q'}^{\rm(E)}(t')\right)^{{1\over q'}-q'}\frac{\partial}{\partial
v}\Big(p_{q'}^{\rm(E)}(v,\,t')\Big)^{1\over q'}\right] \ ,\\
\end{eqnarray}
i.e., concerning self-similar solutions, both the NFPEs (\ref{FP1})
and (\ref{FPNe}) describe the same kinetic process when considered
in the own time-scale $t$
and $t'$, respectively. \\
In figure 1, we plot the relevant quantities $U_q^{(1)}(t)$ [resp.
$U_{q'}^{\rm(E)}(t')]$, $\bar Z^{(1)}_{2-q}(t)$ [resp. $\bar
Z^{\rm(E)}_{q'}(t')]$, $\beta^{(1)}_{2-q}(t)$ [resp.
$\beta^{\rm(E)}_{q'}(t')]$ and $p_q^{(1)}(0,\,t)$ [resp.
$p_{q'}^{\rm(E)}(0,\,t')]$, for the case $q=1.4$; $D^{(1)}=0.2$ and
$D^{\rm (E)}=1$. In all panels the full line depicts the evolution
of these quantities in the modified 1st-formalism v.s. time $t$,
whilst the dashed line depicts the evolution of the same quantities
in the escort-formalism v.s. the time $t$. Finally, the dotted line
depicts the evolution of all these quantities in the
escort-formalism v.s. the transformed time $t'$. We can see that all
the functions in the escort-formalism, referred to the time $t'$,
coincide with the corresponding ones in the modified 1st-formalism,
referred to the time $t$ for all time\footnote{Note that in the
figure we have slightly shifted, by hand, the dotted curves for the
sake of
presentation.}.\\
\begin{figure}[h]
\hspace{-10mm}
  \resizebox{160mm}{!}{\includegraphics{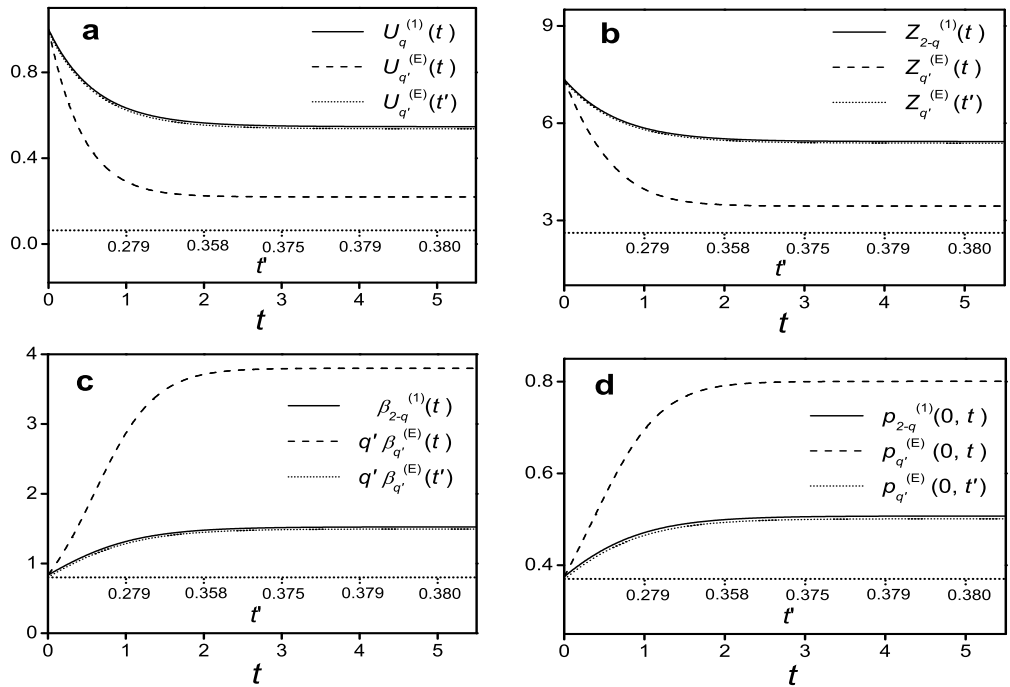}}
  \caption{Plots, in arbitrary unity, of: a)
$U_q^{(1)}(t)$ [resp. $U_{q'}^{\rm(E)}(t')]$, b) $\bar
Z^{(1)}_{2-q}(t)$ [resp. $\bar Z^{\rm(E)}_{q'}(t')]$, c)
$\beta^{(1)}_{2-q}(t)$ [resp. $\beta^{\rm(E)}_{q'}(t')]$, d)
$p_q^{(1)}(0,\,t)$ [resp. $p_{q'}^{\rm(E)}(0,\,t')]$. Full line:
modified 1st-formalism v.s. time $t$; dashed line: escort-formalism
v.s. time $t$; dotted line: escort-formalism v.s. time $t'$.}
\end{figure}

As known, some concepts like irreversibility of the process
described by the Fokker-Planck equation or the $H$-theorem ensuring
uniqueness and stability of equilibrium distribution, can be well
described by means of the Lyapunov function. Existence of a Lyapunov
function for Eqs. (\ref{FP1}) and (\ref{FPNe}) in connection with
the $H$-theorem has been investigated in \cite{Shiino1} and more in
general, for an arbitrary NFPE, in \cite{Frank2}. In the following,
we make some considerations concerning the ``correspondence'' among
the Lyapunov functionals ${\mathcal L}^{(1)}_q[p]$ and ${\mathcal
L}^{\rm(E)}_{q'}[p]$ arising in the two
formalisms.\\
Firstly, we observe that the both functionals ${\mathcal L}^{\rm
i}_q[p]$ given in Eqs. (\ref{Lyapunov1}) and (\ref{Lyapunov3}),
where i=(1) or (E), are actually Lyapunov functionals for the
kinetic equations (\ref{FP1}) and (\ref{FPNe}), respectively, since
they fulfill the relation
\begin{equation}
\frac{d\,{\mathcal L}^{\rm i}_q}{d\,t}=-2\,U^{\rm
i}(t)\,\Big(f_q^{\rm i}(t)\Big)^2\leq0 \ , \label{3}
\end{equation}
where equality holds at equilibrium ($f^{\rm i}_q(t_{\rm e})=0$).
According to Eq. (\ref{en}) we have
\begin{equation}
\Big(f_{q'}^{\rm(E)}(t')\Big)^{-2}\,\frac{d \,{\mathcal
L}_{q'}^{\rm(E)}}{d\,t'}=\Big(f_q^{(1)}(t)\Big)^{-2}\,\frac{d
\,{\mathcal L}_q^{(1)}}{d\,t} \ ,
\end{equation}
which relates the variation in time of the Lyapunov functions in both formalisms. \\
By recalling Eqs. (\ref{eq0}) and (\ref{eq2}), with $X^{\rm
i}_q(t)=U^{\rm i}_q(t)$, we can rewrite Eq. (\ref{3}) in the form
\begin{equation}
\frac{d\,{\mathcal L}^{\rm i}_q}{d\,t}=\frac{d\, U^{\rm
i}_q}{d\,t}\,f_q^{\rm i}(t) \ ,
\end{equation}
so that
\begin{equation}
\frac{d\,{\mathcal L}^{\rm i}_q}{d U^{\rm i}_q}=f_q^{\rm i}(t) \ ,
\end{equation}
i.e., the functions $f_q^{\rm i}(t)$, responsible for the different
time-scale in the two formalisms, originate from the
variation of the Lyapunov function with respect to the mean energy.\\
Finally, let us remind that the distributions $p^{(3)}_{q'}(v,\,t')$
and $p_{q'}^{\rm(E)}(v,\,t')$ are closely related to each other
according to Eq. (\ref{p1-p3}). By inserting this transformation
into Eq. (\ref{FPNe}) we obtain the following kinetic equation for
the distribution $p^{(3)}_{q'}(v,\,t)$
\begin{eqnarray}
\nonumber\hspace{-6mm}\frac{\partial}{\partial\,t'}
p^{(3)}_{q'}(v,\,t')={\partial\over\partial\,v}\left[v\,p^{(3)}_{q'}(v,\,t')+\frac{D^{(1)}}{2-q'}\,
\left(\bar Z_{q'}^{(3)}\right)^{2\,(1-q')}\,
{\partial\over\partial\,v}\Big(p_{q'}^{(3)}(v,\,t')\Big)^{2-q'}\right]
\ ,\\\label{FP33}
\end{eqnarray}
where we assumed $D^{(1)}\equiv D^{\rm(E)}$. By means of the
relation
\begin{equation}
D^{(3)}(t')={D^{(1)}\over2-q'}\, \Big(\bar
Z_{q'}^{(3)}(t')\Big)^{2(1-q')} \ ,\label{dd}
\end{equation}
equation (\ref{FP33}) assumes the same expression of Eq.
(\ref{FP1}), but with a time dependent diffusion coefficient.\\ It is worthy to observe that, contrarily to the NFPEs (\ref{FP1}) and (\ref{FPNe}), Eq. (\ref{FP33}) does not follow from Eqs. (\ref{FP}) and (\ref{j}) and it has been derived just on the basis of the ``correspondence'' between self-similar solutions in the
different formalism. From a formal point of view, we can state that kinetic evolution of self-similar solutions, both in the $(2-q)$-formalism and in the
3rd-formalism, are described by the same NFPE with a constant
diffusion coefficient $D^{(1)}$ in the former case and a time
dependent diffusion coefficient $D^{(3)}(t')$ in the later case (or
vice versa). Remarkably, since $1/D^{(1)}=\beta^{(1)}$ and at
equilibrium $1/D^{(3)}(t_{\rm e}) = \beta^{(3)}$, Eq. (\ref{dd})
reduces to Eq.
(\ref{beta}) with $q$ replaced by $q'$.\\
By following the same steps described at the begin of this section,
by replacing the relation
\begin{equation}
\frac{\partial}{\partial v}\left(p^{(3)}_q(v,\,t)\right)^{2-q}
=-v\,(2-q)\,\beta^{(3)}(t)\,p^{(3)}_q(v,\,t) \ ,
\end{equation}
in Eq. (\ref{FP33}), it can be rewritten in the form
\begin{eqnarray}
\nonumber &&\frac{\partial}{\partial\,t''}
p^{(3)}_{q'}(v,\,t'')=\Bigg(1-D^{(3)}\,\beta^{(3)}(t'')\,\left(\bar
Z_{q'}^{(3)}(t'')\right)^{2\,(1-q')}\Bigg)\,\frac{\partial}{\partial
v}\left(v\,p^{(3)}_{q'}(v,\, t'')\right) \ .\\
\end{eqnarray}
In this way, by introducing the (time-evolution) function
\begin{equation}
f_{q'}^{(3)}(t'')=1-D^{(3)}\,\beta^{(3)}(t'')\,\left(\bar
Z_{q'}^{(3)}(t'')\right)^{2\,(1-q')} \ ,
\end{equation}
we can establish a ``correspondence'' between self-similar solutions
in the modified 1st-formalism and those of the 3rd-formalism, in the
sense that, according to the transformation
\begin{equation}
p_{q'}^{(3)}(v,\,t'')=\left(\bar
Z_{2-q}^{(1)}(t)\right)^{1-q}\,\left(p_q^{(1)}(v,\,t)\right)^{2-q} \
,
\end{equation}
and
\begin{equation}
\frac{dt''}{dt}=\frac{f_q^{(1)}(t)}{f_{q'}^{(3)}(t'')} \ ,
\end{equation}
the NFPE (\ref{FP33}) turns into Eq. (\ref{FP1}).


\sect{Conclusions}

In the present letter, we have compared self-similar solutions of
some NFPEs obtained from different generalized entropies. For the
sake of simplicity we assumed a quadratic form of the energy density
but extension to a more general expression
$\epsilon(v)=k_0+k_1\,v+k_2\,v^2$, with $k_i>0$, is
straightforward.\\ Our results can be summarized in the following two main points.\\
1) According to Eqs. (\ref{beta}) and (\ref{u}) we have shown that
solutions $p(v,\,t)$ of the NFPE (\ref{FP1}), derived in the
$(2-q)$-formalism, can be written in the form $p_q^{(1)}(v,\,t)$
obtained in the modified 1st-formalism or in the form
$p_q^{(3)}(v,\,t)$ obtained in the 3rd-formalism.\\ 2) We have shown
that the solution $p_q^{(1)}(v,\,t)$ of the NFPE (\ref{FP1})
``corresponds'' to the solution $p_{q'}^{\rm(E)}(v,\,t)$ of the NFPE
(\ref{FPNe}) derived in the escort-formalism, where $q=2-1/q^\prime$. Remark that this relation embodies both dualities $q\to2-q$ and
$q\to 1/q$. These two transformations appear recurrently,
alone or combined, in the framework of the generalized statistical
mechanics based on the Tsallis entropy
\cite{Tsallis6,Robledo1,Robledo2,Burlaga,Tsallis7,Tsallis8}.\\ The
``correspondence'' $p_q^{(1)}(v,\,t)=p_{q'}^{\rm (E)}(v,\,t')$ is
established by means of different time-scale, stated through Eq.
(\ref{tt}). As a consequence, it has been shown that the
(self-similar) solutions, obtained in the modified 1st-formalism and
the ones obtained in the 3rd-formalism, obey to similar kinetic
equations where the diffusion coefficients are related according to
Eq. (\ref{dd}).\\
This paper makes some progress about the still discussed question of
the equivalence between the various formalisms, introduced in the
generalized statistical mechanics. We have shown that the
formulation of the theory centered around the escort mean values
(\ref{esc}), whose physical meaning is still unclear at all, are in
any way valid, being related to the same results obtained by
employing the orthodox mean value definition (\ref{lin}). This
encourages the use of the escort probabilities which, in some cases,
expedites the
derivation of several results concerning this theory.\\
On a general basis, we have verified that concerning the class of
self-similar solutions, there are some different nonlinear diffusive
equations which describe the same kinetic process when it is
observed in the appropriate time-scale. The time-scale is derived by
means of the Lyapunov function which generates the corresponding
kinetics evolution equation. In this way, by modifying the expression
of the Lyapunov function, we can produce a family of nonlinear
Fokker-Planck equations belonging to the same equivalent class, in
the sense discussed in this work. It should be interesting to verify
the existence of these equivalences also for more general kind of
solutions.


\app\sect{A}

We summarize, for the sake of comparison, the expressions of some physically relevant quantities like the entropy, the mean energy constraint and the distribution for the different formalisms discussed in this paper.\\ All these quantities are related by means of the following variational principle
\begin{equation}
{\delta\over\delta p}\Big(S[p]-\gamma\,\Phi_0[p]-\beta\,\Phi_1[p]\Big)=0 \ ,
\end{equation}
where $\gamma$ and $\beta$ are the Lagrange multipliers related to the constraints (given through $\Phi_0[p]$ and $\Phi_1[p]$) of the normalization and that of the mean energy, respectively.\\

a) ($2-q$)-formalism.
\begin{eqnarray}
&&S_{2-q}[p]={1\over q-1}\int \left[p(v)^{2-q}-p(v)\right]\,dv \ ,\\ &&\Phi_1[p]=\int\epsilon(v)\,p(v)\,dv-U_q \ ,\\
&&p(v)=\alpha\,\exp_q\Big(-\gamma-\beta\,\epsilon(v)\Big)
\ .
\end{eqnarray}

b) Modified 1st-formalism.
\begin{eqnarray}
&&S_q[p]={1\over 1-q}\int\left[(p^{(1)}(v))^q-p^{(1)}(v)\right]\,dv \ ,\\ &&\Phi_1[p]=\int\epsilon(v)\,p^{(1)}(v)\,dv-U_q^{(1)} \ ,\\
&&p^{(1)}(v)={1\over\bar
Z_q^{(1)}}\,\exp_{2-q}\Big(-\beta_q^{(1)}\,\Big(\epsilon(v)-U^{(1)}_q\Big)\Big)
\ ,\\
&&\beta^{(1)}_q={\beta\over q}\,\left(\bar Z_q^{(1)}\right)^{q-1} \ ,\\
&&\bar
Z_q^{(1)}=q^{1/(q-1)}\,\exp_q(\gamma+\beta\,U^{(1)}_q) \ .
\end{eqnarray}

c) 2nd-formalism.
\begin{eqnarray}
&&S_q[p]={1\over1-q}\int \left[(p^{(2)}(v))^q-p^{(2)}(v)\right]\,dv \ ,\\ &&\Phi_1[p]=\int\epsilon(v)\,(p^{(2)}(v))^q\,dv-U_q^{(2)} \ ,\\
&&p^{(2)}(v)={1\over
Z_q^{(2)}}\,\exp_q\Big(-\beta\,\epsilon(v)\Big)
\ ,\\
&&Z_q^{(2)}=q^{1/(q-1)}\,\exp_q(\gamma) \ .
\end{eqnarray}

d) 3rd-formalism.
\begin{eqnarray}
&&S_q[p]={1\over1-q}\int \left[(p^{(3)}(v))^q-p^{(3)}(v)\right]\,dv \ ,\\ &&\Phi_1[p]={\int\epsilon(v)\,(p^{(3)}(v))^q\,dv\over\int (p^{(3)}(v))^q\,dv}-U_q^{(3)} \ ,\\
&&p^{(3)}(v)={1\over\bar
Z_q^{(3)}}\,\exp_q\Big(-\beta_q^{(3)}\,\Big(\epsilon(v)-U^{(3)}_q\Big)\Big)
\ ,\\
&&\beta^{(3)}_q=\beta\,\left(\bar
Z_q^{(3)}\right)^{q-1} \ ,\\
&&\bar
Z_q^{(3)}=q^{1/(q-1)}\,\exp_q(\gamma) \ .
\end{eqnarray}

e) OLM-formalism.
\begin{eqnarray}
&&S_q[p]={1\over1-q}\int \left[(p^{(\rm OLM)}(v))^q-p^{(\rm OLM)}(v)\right]\,dv \ ,\\ &&\Phi_1[p]={\int\epsilon(v)\,(p^{\rm(OLM)}(v))^q\,dv-U_q^{\rm(OLM)}\int (p^{\rm(OLM)}(v))^q\,dv} \ ,\\
&&p^{\rm(OLM)}(v)={1\over\bar
Z_q^{\rm(OLM)}}\,\exp_q\Big(-\beta\,\Big(\epsilon(v)-U^{\rm(OLM)}_q\Big)\Big)
\ ,\\
&&\bar
Z_q^{\rm(OLM)}=q^{1/(q-1)}\,\exp_q(\gamma) \ .
\end{eqnarray}

f) Escort-formalism.
\begin{eqnarray}
&&S_q^{\rm (E)}[p]={1\over1-q}\left\{{\int p^{\rm(E)}(v)\,dv\over\left[\int\left(p^{\rm (E)}(v)\right)^{1/q}\,dv\right]^q}-1\right\} \ ,\\ &&\Phi_1[p]=\int\epsilon(v)\,p^{\rm(E)}(v)\,dv-U_q^{\rm(E)} \ ,\\
&&p^{\rm(E)}(v)={1\over \bar
Z_q^{(E)}}\,\left[\exp_q\left(-\beta_q^{\rm(E)}
\Big(\epsilon(v)-U^{\rm (E)}_q\Big)\right)\right]^q \
,\\
&&\beta_q^{\rm(E)}=\beta\,\left(\bar
Z_q^{\rm(E)}\right)^{q-1} \ ,\\
&&\gamma=-\beta\,U^{\rm(E)}_q \ .
\end{eqnarray}




\end{document}